\documentclass[12pt]{article}
\usepackage{amssymb}
\usepackage{amsmath}
\usepackage[space]{cite}
\usepackage{algorithm}
\usepackage{bbm}
\usepackage[noend]{algpseudocode}
\usepackage[normalem]{ulem}
\usepackage{url}
\usepackage{graphicx}
\usepackage{adjustbox}
\usepackage{caption}
\usepackage{subcaption}
\usepackage{enumitem}
\usepackage{xcolor}

\usepackage{MnSymbol}

\usepackage{cases}

\numberwithin{equation}{section}

\newcommand{\non}{\nonumber}

\newcommand{\cut}[1]{\ifmmode\text{\textcolor{red}{\sout{\ensuremath{#1}}}}\else\textcolor{red}{\sout{#1}}\fi}

\newcommand{\RN}[1]{%
  \textup{\uppercase\expandafter{\romannumeral#1}}%
}

\begin{document}

\begin{titlepage}
\vspace{.5in}
\begin{center}

{\LARGE Estimating Bethe roots with VQE}\\
\vspace{1in}

\large David Raveh\footnote{\tt 
dxr921@miami.edu}${}^{, 3}$\quad and \quad 
Rafael I. Nepomechie\footnote{\tt 
nepomechie@miami.edu}${}^{,}$\footnote{Corresponding author}\\[0.2in] 
Physics Department, PO Box 248046\\[0.2in] 
University of Miami, Coral Gables, FL 33124 USA

\end{center}

\vspace{.5in}

\begin{abstract}
Bethe equations, whose solutions determine exact eigenvalues and eigenstates of corresponding integrable Hamiltonians, are generally hard to solve.
We implement a Variational Quantum Eigensolver (VQE) approach to estimating Bethe roots of the spin-1/2 XXZ quantum spin chain, by using
Bethe states as trial states, and treating Bethe roots as variational parameters. In numerical simulations of systems of size up to 6, we obtain estimates for Bethe roots corresponding to both ground states and excited states with up to 5 down-spins, for both the closed and open XXZ chains. This approach is not limited to real Bethe roots.
\end{abstract}

\end{titlepage}

\setcounter{footnote}{0}

\section{Introduction}\label{sec:intro}

It is a remarkable fact that certain one-dimensional
quantum many-body models, such as the spin-1/2 XXZ quantum spin chain, are quantum integrable, and can therefore be ``exactly solved'' by Bethe ansatz. (For reviews, see e.g. \cite{Baxter1982, Gaudin:1983, Korepin:1993, Giamarchi2004}.)
The caveat is that the exact eigenvalues and eigenstates of such integrable Hamiltonians are expressed in terms of solutions 
(so-called Bethe roots) of a system of equations called \emph{Bethe equations}; and the latter are generally hard to solve. Indeed, in addition to the venerable approach of Newton's method,
various classical methods have been developed to solve Bethe equations (see e.g. \cite{Hagemans:2007, Hao:2013jqa, Marboe:2016yyn}); however, except when all the Bethe roots are real (in which case the Bethe equations can be efficiently solved by iteration), these methods
generally work only for small system size $L$.

It is natural to ask whether quantum computers can provide an advantage for determining Bethe roots. A first attempt to address this question for the closed XXZ spin chain was made in \cite{Nepomechie:2020}, where it was suggested to use a Variational Quantum Eigensolver approach
(see e.g. \cite{Tilly:2021jem} and references therein) with Bethe states as trial states, treating Bethe roots as variational parameters. However, this effort did not go beyond $M=1$ (where $M$ is the number of down-spins), due to the difficulty of preparing the exact Bethe states. A probabilistic algorithm for preparing these Bethe states was then developed in \cite{VanDyke:2021kvq}, and it was subsequently generalized for the $U(1)$-invariant open XXZ chain in \cite{VanDyke:2021nuz}. However, these algorithms were limited to real Bethe roots; and the success probability was shown to decrease super-exponentially with $M$ \cite{Li:2022czv}. A deterministic algorithm for preparing Bethe states for the closed XXZ chain was developed in \cite{Sopena:2022ntq}; however, it required performing QR decompositions of matrices whose sizes scale exponentially with $M$. Analytic formulae for the unitaries of \cite{Sopena:2022ntq} were proposed in \cite{Ruiz:2023rew}. An explicit deterministic algorithm for preparing Bethe states for both the closed and open XXZ chains, for general values of $L$ and $M$, was recently formulated in \cite{Raveh:2024llj}, see also \cite{Mao:2024hfg}.

In the present work, we revisit the idea \cite{Nepomechie:2020} of estimating Bethe roots by treating them as variational parameters in a VQE setup that uses Bethe states -- prepared now with the new algorithm \cite{Raveh:2024llj} -- as trial states. In numerical simulations
of systems of size up to $L=6$, we indeed obtain estimates for Bethe roots (some of which are complex) corresponding to both ground states and excited states with up to $M=5$ down-spins, for both the closed and open XXZ chain. 

The outline of the remainder of the paper is as follows. In Sec. \ref{sec:basics} we briefly review the closed and open XXZ chains and their corresponding Bethe equations, thereby also laying out our notations. In Sec. \ref{sec:GS}, we report results from our simulations for ground-state
Bethe roots, obtained by minimizing the expectation value of the Hamiltonian using Bethe states as trial states. In Sec. \ref{sec:ES}, we 
report our results for excited-state Bethe roots, obtained by minimizing the Hamiltonian variance \cite{Zhang:2020bhg} using Bethe states as trial states. We end in Sec. \ref{sec:discuss} with a brief discussion of our results. Code in Qiskit \cite{Qiskit} implementing these computations is available as Supplementary Material.

\section{The models and their Bethe equations}\label{sec:basics}

We briefly review here the models and their corresponding Bethe equations, thereby also establishing our notations.

\subsection{Closed chain}\label{sec:CBasics}

For the closed spin-1/2 XXZ quantum spin chain 
of length $L$ with periodic boundary conditions, we define the Hamiltonian
\begin{equation}
    \mathcal{H} = \tfrac{1}{4}\sum_{n=1}^L 
    \left(\sigma^x_n \sigma^x_{n+1} + \sigma^y_n \sigma^y_{n+1} + \Delta\, \sigma^z_n \sigma^z_{n+1} \right) \,, \qquad \vec{\sigma}_{L+1} = \vec{\sigma}_1 \,,
\label{Hclosed}
\end{equation}
where as usual $\sigma^x_n, \sigma^y_n, \sigma^z_n$ are Pauli matrices at site $n$, and $\Delta$ is the anisotropy parameter. The Hamiltonian has the $U(1)$ symmetry
\begin{equation}
    \left[  \mathcal{H} \,, S^z \right] = 0 \,, \qquad S^z = \sum_{n=1}^L \tfrac{1}{2}\sigma^z_n \,.
\end{equation}

The corresponding Bethe equations for the so-called Bethe roots $\vec{k} = (k_1, \ldots, k_M)$ are given by (see e.g. \cite{Baxter1982, Gaudin:1983}) \footnote{Bethe roots are often expressed instead in terms of variables $u_j$, which are related to $k_j$ by $e^{i k_j} = \frac{\sinh(u_j + \eta/2)}{\sinh(u_j - \eta/2)}$, where $\Delta = \cosh \eta$. The Bethe equations \eqref{BEclosed} then take the form
\begin{equation}
    \left( \frac{\sinh(u_j + \frac{\eta}{2})}{\sinh(u_j - \frac{\eta}{2})}\right)^L = \prod_{l=1; l\ne j}^M 
    \frac{\sinh(u_j -u_l + \eta)}{\sinh(u_j -u_l - \eta)} \,, \qquad j = 1, \ldots, M \,. \non
\end{equation} \label{footnote:BEclosed}}
\begin{equation}
    e^{i k_j L} =  \prod_{l=1; l\ne j}^M \left(- \frac{s(k_l, k_j)}{s(k_j, k_l)} \right) \,, \qquad j = 1, \ldots, M \,,
    \label{BEclosed}
\end{equation}
where
\begin{equation}
s(k, k') = 1 - 2 \Delta e^{i k'} + e^{i(k+k')} \,.
\label{s(k,k')}
\end{equation}
The order of Bethe roots $k_i$ in $\vec{k}$ does not matter, since the
Bethe equations are invariant under the interchange of any pair
$(k_j\,, k_l)$.  Note also that the Bethe equations remain invariant
under a $2\pi$ shift of any Bethe root, $k_j \mapsto k_j + 2\pi$;
hence, we can restrict $-\pi < \Re e(k_j) \le \pi$.  For $M=1$, the
empty product on the right-hand-side of \eqref{BEclosed} is understood
to give 1; the same applies to the empty products for $M=1$ in Eqs.
\eqref{Acoeffclosed}, \eqref{BEopen}, \eqref{Acoeffopen}, etc.  below.
 
Indeed, the exact eigenstates of the Hamiltonian \eqref{Hclosed}
are given (up to a normalization factor) by the Bethe states
\begin{equation}
    |B^L_M (\vec{k})\rangle= \sum_{w\in P(L,M)}f(w, \vec{k})\,|w\rangle\,,
    \label{BLM}
\end{equation}
where the sum is over the set $P(L,M)$ of all permutations $w$ with $L-M$ zeros (up-spins) and $M$ ones (down-spins). Furthermore, $f(w,\vec{k})$
is given by
\begin{equation}
    f(w, \vec{k}) = \sum_{\sigma\in S_M} \varepsilon(\sigma)\, A(k_{\sigma(1)}, \cdots, k_{\sigma(M)})\, 
    e^{i \sum_{j=1}^M k_{\sigma(j)} x_j} \,,
\label{fXXZclosed}
\end{equation} 
where the sum is over the set $S_M$ of all permutations $\sigma$ of $\{1,\dots,M\}$, and $\varepsilon(\sigma)=\pm 1$ denotes the sign of the permutation $\sigma$. Moreover,
\begin{equation}
A(k_1, \ldots, k_M) = \prod_{1\le j<l\le M}  s(k_l, k_j) \,, 
\label{Acoeffclosed}
\end{equation} 
and the $x_j \in \{1, \ldots,L\} $ in \eqref{fXXZclosed} are the positions of the 1's in the argument $w$ of $f(w, \vec{k})$. 
The corresponding eigenvalues of $\mathcal{H}$ and $S^z$ are respectively given by
\begin{equation}
    E = \sum_{j=1}^M \cos(k_j) + \left(\frac{L}{4} - M\right) \Delta \,, \qquad S^z = \frac{L}{2} - M \,.
\label{simeigenvaluesclosed}
\end{equation}

The Hamiltonian \eqref{Hclosed} has the charge conjugation (or duality) symmetry
\begin{equation}
    \mathcal{C}\, \mathcal{H}\, \mathcal{C} = \mathcal{H} \,,
    \qquad \mathcal{C} = \left(\sigma^x \right)^{\otimes L} \,, 
    \qquad \mathcal{C}^2 = 1 \,.
\label{duality}
\end{equation}
The fact $\mathcal{C}\, S^z \, \mathcal{C} = - S^z$ implies a 
two-fold degeneracy for states with $S^z \ne 0$, and allows us to restrict $M\le \lfloor L/2 \rfloor$.\footnote{If $|B^L_M(\vec k)\rangle$ is an eigenstate with $M$ Bethe roots, then $\mathcal{C}\,|B^L_M(\vec k)\rangle$ is an eigenstate with the same energy and opposite value of $S^z$, but with
$L-M$ Bethe roots.} (For $M=0$, the Bethe state is proportional to the reference state $|0\rangle^{\otimes L}$, where $|0\rangle=\binom{1}{0}$.)

\subsection{Open chain}\label{sec:OBasics}

For the $U(1)$-invariant open spin-1/2 XXZ quantum spin chain 
of length $L$, we define the Hamiltonian
\begin{equation}
    \mathcal{H} = \tfrac{1}{4}\sum_{n=1}^{L-1}
    \left(\sigma^x_n \sigma^x_{n+1} + \sigma^y_n \sigma^y_{n+1} + \Delta\,  \sigma^z_n \sigma^z_{n+1} \right) + \tfrac{1}{4}\left(h\, \sigma^z_1 + h'\, \sigma^z_L \right) \,,
\label{Hopen}
\end{equation}
where again $\Delta$ is the anisotropy parameter, and
$h$ and $h'$ are boundary magnetic fields. 

The corresponding Bethe equations for the Bethe roots 
$\vec{k} = (k_1, \ldots, k_M)$ are given by \cite{Alcaraz:1987uk}\footnote{In terms of the $u$-roots in Footnote \ref{footnote:BEclosed}, the Bethe equations \eqref{BEopen} take the form
\begin{equation}
    \frac{\sinh(u_j - \frac{\eta}{2} + a )}
    {\sinh(u_j + \frac{\eta}{2} - a)}
    \frac{\sinh(u_j - \frac{\eta}{2} + b )}
    {\sinh(u_j + \frac{\eta}{2} - b)}
    \left( \frac{\sinh(u_j + \frac{\eta}{2})}{\sinh(u_j - \frac{\eta}{2})}\right)^{2L} = \prod_{l=1; l\ne j}^M \left(
    \frac{\sinh(u_j -u_l + \eta)}{\sinh(u_j -u_l - \eta)}
    \frac{\sinh(u_j + u_l + \eta)}{\sinh(u_j + u_l - \eta)}\right)\,,  \non
\end{equation}
where $h= \sinh \eta \coth b$ and $h'= \sinh \eta \coth a$.
}
\begin{equation}
\frac{\alpha(k_j)\, \beta(k_j)}{\alpha(-k_j)\, \beta(-k_j)}
= \prod_{l=1; l\ne j}^M \frac{B(-k_j, k_l)}{B(k_j, k_l)} 
\,, \qquad j = 1, \ldots, M \,,
\label{BEopen}
\end{equation}
where 
\begin{equation}
B(k,k') = s(k, k')\, s(k', -k) \,,
\end{equation}  
see \eqref{s(k,k')},
and
\begin{equation}
\alpha(k) = 1 + (h-\Delta)\,e^{-i k} \,, \qquad
\beta(k) = \left[1 + (h'-\Delta)\,e^{-i k} \right] e^{i(L+1) k}\,.  
\end{equation}
The Bethe equations \eqref{BEopen} remain invariant under a $2\pi$ shift of any Bethe root, $k_j \mapsto k_j + 2\pi$, as well as under the reflection $k_j \mapsto -k_j$ ; hence, we can restrict $0 \le \Re e(k_j) \le\pi$. 

Indeed, the exact simultaneous eigenvectors of the Hamiltonian $\mathcal{H}$  \eqref{Hopen} and $S^z$ again have the form 
$|B^L_M (\vec{k})\rangle$ \eqref{BLM}, where $f(w, \vec{k})$ is now given by
\begin{equation}
    f(w, \vec{k}) = \sum_{\sigma\in S_M}
    \sum_{\epsilon_1,\dots,\epsilon_M=\pm 1}
    \varepsilon(\sigma)\,\epsilon_1\dots\epsilon_M\, A(\epsilon_1 k_{\sigma(1)}, \ldots, \epsilon_M k_{\sigma(M)})\, 
    e^{i \sum_{j=1}^M \epsilon_j k_{\sigma(j)} x_j} \,.
\label{fXXZopen}
\end{equation} 
Moreover, now
\begin{equation}
A(k_1, \ldots, k_M) = \prod_{j=1}^{M} \beta(-k_j) 
\prod_{1\le j<l\le M} B(-k_j, k_l)\, e^{-i k_l} \,.
\label{Acoeffopen}
\end{equation}
The $x_j \in \{1, \ldots,L\} $ in \eqref{fXXZopen} are again the positions of the 1's in the argument $w$ of $f(w, \vec{k})$.
The corresponding eigenvalues of $\mathcal{H}$ and $S^z$
are now given by 
\begin{equation}
    E = \sum_{j=1}^M \cos(k_j) + \left(\tfrac{1}{4}(L-1) - M\right) \Delta + \tfrac{1}{4}(h + h')\,, \qquad S^z = \frac{L}{2} - M \,.
\label{simeigenvaluesopen}
\end{equation}
For generic values of the parameters $\Delta$, $h$ and $h'$, the spectrum is non-degenerate, and $M \in \{0 , 1, \ldots, L\}$.
(For $M=L$, the Bethe state is proportional to the dual reference state $|1\rangle^{\otimes L}$, where $|1\rangle=\binom{0}{1}$.)

\section{Ground-state Bethe roots}\label{sec:GS}

In this section, we apply the  hybrid quantum/classical VQE algorithm \cite{Tilly:2021jem}, using Bethe states $|B^L_M (\vec{k})\rangle$ \eqref{BLM} as trial states and treating the Bethe roots $\vec{k}=(k_1, \ldots, k_M)$ as variational parameters, in order to estimate the Bethe roots corresponding to \emph{ground states}. The key point is that, according to the variational theorem, for \emph{arbitrary} values of $\vec k$ we have the inequality
\begin{equation}
\label{expectation}
    E_0\leq\langle B^L_M (\vec{k}) | \mathcal{H} | B^L_M (\vec{k})\rangle\,,
\end{equation}
where the equality holds if and only if $\vec k$ are the Bethe roots corresponding to the ground state.
The standard VQE algorithm begins by randomly selecting initial values $\vec k=\vec{k}^{(0)}$, and seeks to vary $\vec k$ so that the expectation \eqref{expectation} is minimized via the following steps:
\begin{enumerate}
    \item Given $\vec k=\vec{k}^{(i)}$, prepare the state $|B^L_M (\vec{k}^{(i)})\rangle$ on a quantum computer.
    \item Compute the expectation $\langle B^L_M (\vec{k}^{(i)}) | \mathcal{H} | B^L_M (\vec{k}^{(i)})\rangle$ on a quantum computer.
    \item Repeat steps 1-2 for some points $\vec k=\vec{k}^{(i)}+\vec \epsilon$ in an $\vec{\epsilon}$-neighborhood of $\vec{k}^{(i)}$. Use these values to compute with a classical optimizer the numerical gradient of $\langle B^L_M (\vec{k}^{(i)}) | \mathcal{H} | B^L_M (\vec{k}^{(i)})\rangle$ about the point $\vec{k}^{(i)}$, and 
    select $\vec k=\vec{k}^{(i+1)}$ in the direction of steepest descent.
    \item Repeat steps 1-3 with $i\to i+1$ until the expectation converges to the ground-state energy.
\end{enumerate}

 We implement this approach using Qiskit v1.0 \cite{Qiskit}, which provides the necessary functionality for a given choice of
 trial state, Hamiltonian, and optimizer; our code is available as Supplementary Material. We simulate the results rather than using a real quantum device. In step 1, we prepare the state $|B^L_M (\vec{k}^{(i)})\rangle$ using the circuit in\cite{Raveh:2024llj}. In step 2, the expectation $\langle B^L_M (\vec{k}^{(i)}) | \mathcal{H} | B^L_M (\vec{k}^{(i)})\rangle$ is computed in one of two ways: using Qiskit's Statevector simulator, which actually carries out the matrix multiplication (and hence is exact), or using Qiskit's noiseless Aer simulator. The Aer simulator computes the expectation approximately, operating via \emph{shots}, or repeated trials.\footnote{There are several ways to measure the expectation of a Hamiltonian expressed as a sum  $\mathcal{H}=\sum_i a_i\,U_i$ of Pauli strings $U_i$, e.g. $U_i=\sigma_j^x\sigma_{k}^z$ on a quantum computer. In one method, $p_i=\langle B^L_M (\vec{k}) | U_i | B^L_M (\vec{k})\rangle$ can be computed via the Hadamard test, which requires repeated trials. It follows that $\langle B^L_M (\vec{k}) | \mathcal{H} | B^L_M (\vec{k})\rangle=\sum_i a_i\,p_i$.\label{footnote:expectation}} For this algorithm to run on a real device, one would need to implement a shot-based computation, and so the Aer simulator results reflect the result of operating this algorithm on a perfect (noiseless) quantum computer. In step 3, we adopt Qiskit's Cobyla as the classical optimizer. In step 4, we iterate the algorithm no more than 1000 times, although typically convergence occurs far sooner; and as the expectation converges to the ground-state energy, our variational parameters converge to the ground-state Bethe roots. 

We first consider in Sec. \ref{sec:CG} the closed XXZ chain \eqref{Hclosed} with $L$ even in the antiferromagnetic regime $\Delta>0$, where the non-degenerate ground state is known to be described by $M= L/2$ real Bethe roots. Next we consider in Sec. \ref{sec:OG} the open XXZ chain \eqref{Hopen}, choosing values for the boundary fields such that the ground state is no longer described by only real Bethe roots. We then analyze the accuracy of the results in Sec. \ref{sec:error}.

\subsection{Closed chain}\label{sec:CG}

We begin by implementing the VQE approach sketched above to the closed XXZ Hamiltonian \eqref{Hclosed} with $\Delta = 2$, for 
even values of $L$ up to $L=6$.
As already noted, the ground state is known to be described by $M=L/2$ real Bethe roots. 

Our results are displayed in Table \ref{table:CG}. The ``Energy'' is computed by direct diagonalization of the Hamiltonian \eqref{Hclosed}; the ``True roots'' are computed by solving the Bethe equations \eqref{BEclosed} using Newton's method, which is implemented in
Mathematica by {\tt FindRoot};\footnote{Since here the Bethe roots are real, they can be obtained more readily by solving the logarithm of the Bethe equations by iteration.} the ``Statevector roots'' are computed using the Qiskit exact Statevector simulator; and the ``Aer roots'' are computed using the Qiskit noiseless shot-based Aer simulator with 10,000 shots. 

\begin{table}[h!]
\small
\centering
\begin{tabular}{||c c c c c c||} 
 \hline
 $L$ &$M$ &Energy &True roots &Statevector roots &Aer roots\\ [0.5ex] 
 \hline\hline
 2 & 1 & $-2$ & $3.14159$ & $3.1415$ & $3.1487$ \\ 
 4 & 2 & $-2.73205$ & $\pm 1.94553$ & $1.9455 ,  -1.9455$  & $1.9623, -1.9503$ \\
 6 & 3 & $-3.85577$ & $\pm 1.49862, 3.14159$ & $ 1.4986 ,  - 1.4986,3.1416$  & $ 1.5477,-1.4830, 3.1796$\\
 \hline
\end{tabular}
\caption{Ground-state Bethe roots for the closed chain \eqref{Hclosed}}
\label{table:CG}
\end{table}

\subsection{Open chain}\label{sec:OG}

We now consider the open XXZ Hamiltonian \eqref{Hopen} with $\Delta = 1/2$, $h=3$, $h'=3/10$, and values of $L$ up to $L=6$. The ground state is described here by $M= \lceil L/2 \rceil$ Bethe roots; and the Bethe roots are not all real. Thus, for each complex root $k_j=\Re e(k_j)+i \Im m(k_j) $, two real variational parameters are introduced.

Our results are displayed in Table \ref{table:OG}. The columns are computed similarly as in Table \ref{table:CG}.

\begin{table}[h!]
\small
\centering
\begin{tabular}{||c c c c c c ||} 
 \hline
 $L$ & $M$ &Energy &True roots &Statevector roots &Aer roots\\ [0.5ex] 
 \hline\hline
 2 & 1 & $-0.965015$ & $3.14159 + 0.882174 i$ & $3.1415+ 0.8822i$ & $3.0298 + 0.8785i$ \\
 3 & 2 & $-1.49506$ & $3.14159 + 0.908996i, $ & $3.1412 + 0.9071i, $  & $3.1537 + 0.9014i,$\\
 & & & $1.69883$ &$1.6988$ & $ 1.6930$\\
 4 & 2 & $-1.76803$ & $3.14159 + 0.91503i, $ & $3.1412+0.9142i, $  & $3.0944+ 0.9929i, $ \\
 & & &$2.11689$ & $2.1169$ & $2.1473$ \\
 5 & 3 & $-2.22762$ & $3.14159 + 0.916011i, $ & $3.1399 + 0.9144i,$ & $3.3120+1.5000i,$ \\
 & & & $1.49569, 2.31576$ & $1.4958, 2.3157$& $1.5449, 2.3291$\\
 6 & 3 & $-2.53682$ & $3.14159 + 0.916239i, $ &  $3.1419 + 0.9131i, $ & $3.0864 + 0.9053i, $ \\
 & & & $1.82675, 2.47141$ & $1.8266, 2.4712$ & $2.0142, 2.3022$\\
 \hline
\end{tabular}
\caption{Ground-state Bethe roots for the open chain \eqref{Hopen}}
\label{table:OG}
\end{table}

\subsection{Error}\label{sec:error}

The Bethe root values displayed in Tables \ref{table:CG} and \ref{table:OG}
are not of interest per se; rather, we focus on the \emph{error} in our methodology, to verify its viability. 
One source of error is the imperfect classical optimizer, which of course behaves more poorly as $M$ increases; more sophisticated optimizers than Cobyla may exist that can perhaps improve upon this error. Additionally, the finite number of iterations in the VQE account for some error, which can be mitigated by increasing the number of iterations.  An important source of error, unique to the Aer roots, is the error associated with the number of shots used to approximate an expectation value. This error $\epsilon$ scales as (see e.g. \cite{Wecker:2015})
\begin{equation}
    \epsilon\sim\frac{1}{\sqrt{x}}\,,
\label{error}
\end{equation}
where $x$ is the number of shots. Thus, with sufficiently many shots this error can be mitigated up to error $\epsilon$ for fixed $L$; the number of expectation values computed in a single VQE iteration grows with $L$, see Footnote \ref{footnote:expectation}. On a real device, there would of course be additional errors due to various types of noise. 

\begin{figure}[htb]
      \centering
\includegraphics[width=0.6\linewidth]{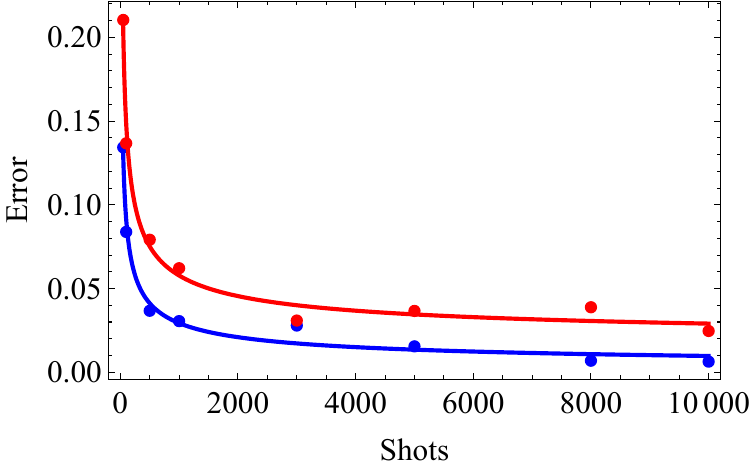}
\caption{The error of the Aer roots $\vec{k}'$ with respect to the True roots $\vec{k}$, defined as $\epsilon=\|\vec k-\vec k'\|/\|\vec k\|$, where $\|\vec k\| = \sqrt{\vec k \cdot \vec k^{*}}$,
for the ground-state Bethe roots of the closed (blue) and open (red) chains when $L=4$, with curves fitted to $\epsilon=a/\sqrt{x}+b$ \eqref{error}. }
\label{fig:error}
\end{figure}

We observe in Tables \ref{table:CG} and \ref{table:OG} that the Statevector results, which represents a VQE with zero shot-based error, are in excellent agreement with the true results. The Aer results, which represent the result of running the VQE algorithm on a noiseless quantum computer, are in fair agreement with the true results; increasing the number of shots improves the accuracy according to \eqref{error}, see Fig. \ref{fig:error}.

\section{Excited-state Bethe roots}\label{sec:ES}

We turn now to the more challenging problem of estimating Bethe roots corresponding to \emph{excited states}. To this end, we proceed similarly as before in Sec. \ref{sec:GS}, except rather than minimizing the expectation of the Hamiltonian $\langle B^L_M (\vec{k}) | \mathcal{H} | B^L_M (\vec{k})\rangle$, we instead
minimize the non-negative \emph{variance} of the Hamiltonian
\begin{equation}    
0\leq\langle \left(\mathcal{H} - \langle \mathcal{H} \rangle \right)^2 \rangle =
\langle B^L_M (\vec{k}) | \mathcal{H}^2 | B^L_M (\vec{k})\rangle - \langle B^L_M (\vec{k}) | \mathcal{H} | B^L_M (\vec{k})\rangle^2\,, 
\end{equation}
where the equality holds for any exact eigenstate, see e.g. \cite{Zhang:2020bhg}. 

It should be noted that this method makes it equally likely to obtain the Bethe roots of any particular eigenstate; that is, there is no bias in favor of low-energy eigenstates. Thus, obtaining the Bethe roots corresponding to a particular eigenstate requires randomly selecting initial parameter values $\vec{k}^{(0)}$ until the output roots correspond to the desired eigenstate.

We first treat the closed chain in Sec. \ref{sec:CE}, followed by the open chain in Sec. \ref{sec:OE}.

\subsection{Closed chain}\label{sec:CE}

We consider the closed XXZ Hamiltonian \eqref{Hclosed} again with $\Delta = 2$, and values of $L$ up to $L=6$. Charge-conjugation invariance \eqref{duality} allows us to restrict to $M\le \lfloor L/2 \rfloor$. Since complex roots of the closed-chain Bethe equations \eqref{BEclosed} always appear in conjugate pairs, we search for each such complex pair of Bethe roots using only two real variational parameters.
As $L$ increases, the number of Bethe states rapidly increases; we therefore present only a representative sample.

Our results are displayed in Table \ref{table:CE}.  The columns are computed similarly as in Table \ref{table:CG}. As for the ground-state Bethe roots, the Statevector results for excited-state Bethe roots are in excellent agreement with the true results, which confirms the viability of this variance-minimization approach. The Aer results are again in fair agreement with the true results; the error behaves similarly as described in Section \ref{sec:error}.

We note that we do not find Bethe roots $\vec{k}$ that include $k=\pm i \infty$ (or $u= \mp \eta/2$ in terms of the $u$-roots in Footnote \ref{footnote:BEclosed}), which are called ``physical singular solutions'' in \cite{Hao:2013jqa, Nepomechie:2013mua}. Indeed, we do not expect to obtain such Bethe roots in this VQE approach, since a delicate limiting procedure is required to obtain the corresponding Bethe states, see e.g. \cite{Nepomechie:2013mua} for the case $\Delta=1$.

\begin{table}[h!]
\small
\centering
\begin{tabular}{||c c c c c c ||} 
 \hline
 $L$ & $M$ &Energy &True roots &Statevector roots &Aer roots\\ [0.5ex] 
 \hline\hline
 2 & 1 & $0$ & $0$ & $0$ & $0.0019$ \\
 3 & 1 & $-1$ & $2.0944$ & $2.0943$ & $2.0938$ \\
 4 & 2 & $0.732051$ & $\pm 0.831443i$ & $\pm 0.8314i$ & $\pm 0.8557i$ \\
 5 & 2 & $0.716341$ & $0.628319 \pm  0.835459i$ & $0.6276 \pm 0.8349i$ & $0.6263 \pm 0.8858i$ \\
 6 & 2 & $-1.75395$ & $1.37766, 2.81114$ & $1.3776, 2.8109$ & $1.3998, 2.8293$ \\
 6 & 3 & $1.18614$ & $0.244998 \pm 1.41247i,$ & $0.2451 \pm 1.4120i, $ & $0.2451 \pm 1.3341i,$ \\
 &&& $1.6044$& $1.6023$ & $1.3152$\\ 

 \hline
\end{tabular}
\caption{Excited-state Bethe roots for the closed chain \eqref{Hclosed}}
\label{table:CE}
\end{table}

\subsection{Open chain}\label{sec:OE}

We now consider the open XXZ Hamiltonian \eqref{Hopen} again with $\Delta = 1/2$, $h=3$, $h'=3/10$, and values of $L$ up to $L=6$. 
Since this model does not have charge-conjugation invariance, we consider values of $M$ up to $M=L-1$.
As for the closed chain, the number of Bethe states increases rapidly with $L$, so we present only a representative sample.

Our results are displayed in Table \ref{table:OE}.  The columns are computed similarly as in Table \ref{table:CG}. Once again, we find excellent agreement 
between the Statevector and true results, and fair agreement between the Aer and true results, similar to the description in Section \ref{sec:error}.

Bethe roots $\vec{k}$ that include $k=\pi$ or
$k= -i \infty$ (that is, $u=0$ or $u= \eta/2$, respectively, in terms of the $u$-roots in Footnote \ref{footnote:BEclosed}) do not appear in Bethe states of the open chain, see e.g. \cite{Fendley:1994cz, Gainutdinov:2015vba}. Hence, in principle, it should be possible to obtain all open-chain Bethe roots by this VQE approach.

\begin{table}[h!]
\small
\centering
\begin{tabular}{||c c c c c c ||} 
 \hline
 $L$ & $M$ &Energy &True roots &Statevector roots &Aer roots\\ [0.5ex] 
 \hline\hline
 2 & 1 & $0.715015$ & $1.30258$ & $1.3026$ & $1.3025$ \\
 3 & 1 & $-0.869852$ & $3.14159 + 0.911371i$ & $3.1417 + 0.9113i$ & $3.1683 + 0.8783i$ \\
 4 & 2 & $-0.224189$ & $3.14159 + 0.916221i, $ & $3.1413 + 0.9164i, $ & $3.0311 + 0.9359i, $ \\
 & & & $0.2264i$&$0.2284i$ & $0.2148i$\\
 4 & 3 & $-0.128194$ & $3.14159 + 0.916237i, $ & $3.1416 + 0.9174i,$ & $3.0528 + 0.9138i,$ \\
 & & & $0.93789, 0.245389i$ & $0.9382, 0.2474i$ & $ 0.9046, 0.2146i$\\
 5 & 4 & $-1.61607$ & $3.14159 + 0.916185i,$ & $3.1421 + 0.9027i,$ & $3.0030 + 0.7578i,$ \\
 & & & $0.514675, 1.16211, 2.43263$ & $0.5044, 1.1618, 2.4332$ & $0.4451, 1.0290, 2.4882$\\
 6 & 5 & $0.21968$ & $3.14159 + 0.916291i,$ & $3.1412 + 0.9156i,$ & $3.1218 + 0.7523i,$ \\
 & & & $0.667057, 0.32195i$ & $0.6252, 0.3029i,$ & $0.5055, 0.5615i,$ \\
 & & & $1.12044 \pm 0.160175i$ & $1.1347 \pm 0.1956i$ & $1.0101  \pm 0.2531i$ \\
 \hline
\end{tabular}
\caption{Excited-state Bethe roots for the open chain \eqref{Hopen}}
\label{table:OE}
\end{table}

\section{Discussion}\label{sec:discuss}

We have seen in noiseless simulations that Bethe roots for both ground states and excited states of the XXZ chain can indeed be estimated using a VQE setup with Bethe states as trial states and Bethe roots as variational parameters. This approach is not limited to real Bethe roots, and can therefore be used to look for solutions of the Bethe equations that cannot be easily obtained classically by iteration.

A limitation of this approach, which is also present in the classical approach of solving Bethe equations using Newton's method, is the need for the initial values of the Bethe roots $\vec{k}^{(0)}$ to be ``sufficiently close'' to the true values. As the number of real variational parameters (which is $\mathcal{O}(M)$) increases, this problem is exacerbated. 
A further limitation is the presence of shot noise, which however can be ameliorated by increasing the number of shots, see Figure \ref{fig:error}. On the other hand, an advantage of our VQE approach for ground-state Bethe roots over directly solving Bethe equations is that, in the former approach, the resulting Bethe roots are guaranteed to be ``physical''; that is, they correspond to actual eigenstates of the Hamiltonian. (When directly solving Bethe equations, one often finds ``unphysical'' solutions, such as those containing repeated roots, which do not correspond to actual eigenstates, see e.g. \cite{Hao:2013jqa}.)

The circuit \cite{Raveh:2024llj} that we used to prepare the trial states renders both the classical and quantum parts of this VQE approach expensive. Indeed, for each VQE iteration, the computation of the $f(w,\vec k)$ coefficients for the closed and open chains respectively require $M!$ and $2^M M!$ summations (see Eqs. \eqref{fXXZclosed} and \eqref{fXXZopen}),  making the classical part expensive for large $M$. Moreover, the circuit has size and depth $\mathcal{O}(\binom{L}{M}$) \cite{Raveh:2024llj}, making also the quantum part expensive for large $L$ and $M$. This VQE approach for estimating Bethe roots is therefore not feasible in the near term (the so-called Noisy Intermediate-Scale Quantum era). However, on fault-tolerant hardware, this approach may have an advantage over classical hardware for moderate values of $L$ and $M$. 

It would be interesting to reduce the classical computational cost of the Bethe state circuit, perhaps by implementing a type of algebraic Bethe ansatz \cite{Sopena:2022ntq, Ruiz:2023rew}, instead of the
coordinate Bethe ansatz implemented in \cite{Raveh:2024llj}. It would also be interesting to reduce the size and/or depth of the Bethe circuit, perhaps by exploiting ancillas, measurements, and feedforward operations as in \cite{Piroli:2024ckr} for Dicke states.

\section*{Acknowledgments} 

We thank John Van Dyke for valuable discussions, and Ananda Roy for reading the draft.
We are also grateful to Daniel Nepomechie and Chana Raveh for their help with python. 
RN is supported in part by the National Science Foundation under grant PHY 2310594, and by a Cooper fellowship.

\providecommand{\href}[2]{#2}\begingroup\raggedright\endgroup

\end{document}